# Account of occasional wave breaking in numerical simulations of irregular water waves in the focus of the rogue wave problem

**Alexey Slunyaev[1)] and Anna Kokorina[1)]**

1) Institute of Applied Physics, Nizhny Novgorod, Russia

Alexey Slunyaev (Corresponding author):
ORCID: 0000-0001-7782-2991
Slunyaev@appl.sci-nnov.ru
Postal address: Institute of Applied Physics, Box-120, 46 Ulyanova Street, Nizhny Novgorod, 603950 Russia

Anna Kokorina:
ORCID: 0000-0001-5196-537X
A.Sergeeva@appl.sci-nnov.ru
Postal address: Institute of Applied Physics, Box-120, 46 Ulyanova Street, Nizhny Novgorod, 603950 Russia

## Abstract

The issue of accounting of the wave breaking phenomenon in direct numerical simulations of oceanic waves is discussed. It is emphasized that this problem is crucial for the deterministic description of waves, and also for the dynamical calculation of extreme wave statistical characteristics, such as rogue wave height probability, asymmetry, etc. The conditions for reproducible simulations of irregular steep waves within the High Order Spectral Method for the potential Euler equations are identified. Such non-dissipative simulations are considered as the reference when comparing with the simulations which use two kinds of wave breaking regularization. It is shown that the perturbations caused by the wave breaking attenuation may be noticeable within 20 min of the wave evolution.

## 1. Introduction

The direct numerical simulation of hydrodynamic equations of has become recently an accessible alternative to the simulation of kinetic equations for oceanic waves. In addition to the open possibility of a direct comparison between results of the dynamic and kinetic approaches (e.g., Annekov & Shrira, 2018), the direct numerical simulations are a promising mean to solve new challenging problems, such as reproduction of the rogue wave phenomenon [Kharif et al, 2009]. In contrast to the kinetic equations, the direct solution of the dynamic equations does not employ the assumption of independent wave phases, which is obviously violated in rough sea conditions. Besides, the degree of idealization of the simulated equations may be chosen appropriately to meet the requirements on capturing significant physical effects as well as on the acceptable computation speed. Today simulations of the sea surface areas of $O(10^2)$ km within the 3D potential Euler equations which account for strongly nonlinear effects may be performed with the speed greater than the waves evolve in the real sea [Köllisch et al, 2018]. Consequently, besides the ability of accumulating the statistical data, the direct numerical simulation is considered as a tool of an operational short-term wave forecast for the needs of navigation [van Groessen et al, 2017a,b; Köllisch et al, 2018].

One may look forward to the forthcoming future when even faster approaches for the simulation of the hydrodynamic equations (maybe with reasonable simplifying assumptions) will be invented. A comprehensive overview of existing approaches may be found in the book by Chalikov (2016), see also new reduced equations for hydrodynamics suggested in [Dyachenko et al, 2017; Zakharov et al, 2002]. The conformal mapping allows exceptionally fast simulations of the full potential equations of hydrodynamics, though it is restricted to the degenerative case of collinear waves (a weakly 3D generalization of the approach was suggested by Ruban (2005)).

Thus, in this work we rely on the idea that the development of approaches to simulate the evolution of irregular sea waves in dynamic equations with the purpose of a statistical study is a worthy cause. The time domain of a possible deterministic forecast is obviously limited due to the uncertainty in the initial data and unaccounted physical effects. Besides the horizon a deterministic wave predictability is limited due to the inner properties of the nonlinear equations of hydrodynamics: According to Annenkov & Shrira (2001) – it is up to $O(10^3)$ wave periods for the realistic sea conditions. Hence the deterministic description may have sense for some limited time, say of the order $O(10^2)$ wave periods or about, and beyond this time one may operate in statistical terms only [Cavalery, 2006].

Among the advantages of the direct numerical simulations of wave ensembles (so-called stochastic simulations) one may point fully controllable conditions of the wave evolution, full and accurate wave data (including kinematic characteristics), manageable accuracy of the governing equations, the high performance of modern computers and implemented algorithms, low price of such research.

The main potential disadvantage of any numerical simulations is the adopted idealizations which may cause failure of description of physical effects or also appearance of unphysical spurious effects. Wave breaking is probably the most obvious physical phenomenon which is difficult for the theoretical modeling of oceanic waves. At the same time this effect is obviously essential for the probability of abnormally high (rogue) waves. The nonlinear wave self-modulation (due to the modulational or Benjamin – Feir instability) is a regular effect capable of generation of extreme sea waves and responsible for the increase of the large wave probability [Onorato et al, 2001, 2002]. In the conditions of irregular sea, the waves should be steep to be pumped by the Benjamin – Feir instability: waves with a small steepness $\varepsilon$ are modulationally unstable with respect to very long perturbations ($\sim\varepsilon^{-1}$ wavelengths) and grow too slowly ($\sim\varepsilon^{-2}$ wave periods). When wave trains self-modulate, the wave amplitude and steepness grow, so that trains consisting of 5-10 waves with the initial steepness $\varepsilon = kH/2 \approx 0.1$ (here $k$ is the wavenumber and $H$ is the wave height) can reach the breaking limit within a few tens of wave periods [Slunyaev & Shrira, 2013]. Therefore even sea states with moderate wave steepness may be affected by the wave breaking phenomenon; and this effect is most pronounced in the large wave characteristics, including extreme wave probability.

Today the High-Order Spectral Method (HOSM) [Dommermuth & Yue, 1987; West et al, 1987] is probably the most popular method for solving the 3D Euler equations. It is significant that HOSM is not a fully nonlinear approach; it resolves accurately up to $M+1$ nonlinear wave interactions, where $M$ is a parameter of the approach. As the computational effort is proportional to the order $M$, the simulations are most frequently limited by the lowest order, $M = 3$ (e.g., Toffoli et al, 2008, 2010; Toffoli & Bitner-Gregersen, 2011; Xiao et al, 2013; Bitner-Gregersen et al, 2014; Bitner-Gregersen & Toffoli, 2014; Brennan et al, 2018), which allows consideration of the predominant in deep waters four-wave interactions including the Benjamin – Feir instability. Certainly such codes cannot simulate waves which tend to break, though they are capable of simulation of the dominating processes leading to the evolution of the wave spectra and may be used to some extent for the description of steep wave events in deep waters.

Crucially, even if the wave breaking process hypothetically could be simulated in detail (for example within the Navier – Stokes equations), this computation would be too slow to be used for the stochastic simulation. As a result, in the stochastic simulations of the Euler equations the wave breaking is always regularized following some simplified empirical approaches. Hence we face the following contradiction: To calculate the wave height probability in the most interesting cases of significant nonlinearity, when the direct numerical simulation may show differences from the weakly nonlinear theory, one must use uncertain parameterizations of the wave breaking, which in its turn controls the occurrence of large waves.

Though the ways of introducing the wave breaking attenuation are physically grounded, and the simulations of wave breaking events are sometimes tested versus laboratory measurements in some limited number of cases, it is clear that the way how the wave breaking is handled may distort the dynamics and appearance of extreme waves, including the probability distribution of high waves. An unphysical early breaking/attenuation limits the maximum wave amplification and hence reduces the rogue wave effect. The wave height and wave crest probability distribution functions are probably the most wanted information in the rogue wave story, meanwhile the wave group structure, wave asymmetry, etc. are also essential, and may be affected by the unnatural mechanisms.

In this study we focus on the problem how the method of regularization of the wave breaking phenomenon manifests itself in the simulated fields of irregular sea waves. We use the framework of primitive non-dissipative Euler equations as the reference case which is not affected by the wave breaking attenuation. The employed numerical codes and the two considered methods of wave breaking regularization are described in Sec. 2. The parameters of the simulations such as the resolution it time and space, which provide reasonably accurate results we discuss in Sec. 3. The reference non-dissipative simulation is compared versus the regularized ones in Sec. 4; this comparison is possible until the first breaking event occurs. Such approach forces us to use the sea state conditions when the wave breakings are not numerous. The two simulations using different methods of the wave breaking regularization are compared between each other in Sec. 5 for longer times. The main conclusions are drawn in the end.

## 2. The numerical code and the methods of wave breaking regularization

The own code for the integration of the potential Euler equations of three-dimensional waves over infinitively deep water is used; the simulations were performed on an ordinary personal computer. The system of governing equations [Zakharov, 1968] consists of two surface boundary conditions (1) and (2) at the water surface $z = \eta(x, y, t)$, the Laplace equation (3) in the water bulk $z \le \eta(x, y, t)$, and the decaying condition (4) at a large depth,

$$\frac{\partial \eta}{\partial t} = -\frac{\partial \Phi}{\partial x}\frac{\partial \eta}{\partial x} - \frac{\partial \Phi}{\partial y}\frac{\partial \eta}{\partial y} + \left(1 + \left(\frac{\partial \eta}{\partial x}\right)^2 + \left(\frac{\partial \eta}{\partial y}\right)^2\right)\frac{\partial \varphi}{\partial z}, \qquad z = \eta, \tag{1}$$

$$\frac{\partial \Phi}{\partial t} = -g\eta - \frac{1}{2}\left(\frac{\partial \Phi}{\partial x}\right)^2 - \frac{1}{2}\left(\frac{\partial \Phi}{\partial y}\right)^2 + \frac{1}{2}\left(\frac{\partial \varphi}{\partial z}\right)^2\left[1 + \left(\frac{\partial \eta}{\partial x}\right)^2 + \left(\frac{\partial \eta}{\partial y}\right)^2\right], \qquad z = \eta, \tag{2}$$

$$\frac{\partial^2 \varphi}{\partial x^2} + \frac{\partial^2 \varphi}{\partial y^2} + \frac{\partial^2 \varphi}{\partial z^2} = 0, \qquad z \le \eta, \tag{3}$$

$$\frac{\partial \varphi}{\partial z} \to 0, \qquad z \to -\infty . \tag{4}$$

Here $\varphi(x, y, z, t)$ is the velocity potential, and $\Phi(x, y, t) = \varphi(x, y, z = \eta, t)$ is the surface velocity potential, and $g$ is the acceleration due to gravity. We employ the popular High-Order Spectral Method [West et al, 1987], which uses the decomposition of the velocity potential into the Taylor series in the vicinity of the surface to bring the varying fluid domain beneath the surface to a rectangular box, which helps to avoid solving the Laplace equation at every step of the integration.

The HOSM has been used in a number recent papers dedicated to the 3D simulations of irregular waves (e.g., Toffoli et al, 2008, 2010; Ducrozet et al, 2007, 2017; Toffoli & Bitner-Gregersen, 2011; Xiao et al, 2013; Bitner-Gregersen et al, 2014; Bitner-Gregersen & Toffoli, 2014; Seiffert et al, 2017; Brennan et al, 2018); in particular, an open source code is available for free use [Ducrozet et al, 2016]. Though the Taylor series which approximate the velocity may be of any order $M$ (then the code takes into account properly $M + 1$ wave interactions, higher-order nonlinear effects are accounted for partly), this number is limited due to the requirement of reasonably fast computation, growing size of the Fourier domain to perform de-aliasing, and the inherent small-scale numerical instability. The accuracy of the approach goes down in the situation of steep waves (then the Taylor series converge poorly) and wide spectra (due to very fast attenuation with depth of the vertical structure of short waves on the top of longer waves). Events may occur in the simulations, when waves become too steep and cause a blowup of the code. Such events are conventionally referred to the wave breaking phenomenon, though strictly speaking one can hardly distinguish with confidence the cases of numerical instability and of the instability caused by the physical wave breaking. By default we will associate a blowup of the simulation with a wave breaking event.

Two methods for the integration it time are used in this work: i) the standard Runge-Rutta 4-order method with fixed time step (RK4) and ii) decoupling of the governing equation into the linear ant nonlinear parts (A&RK4). In the second approach the linear part is calculated at each time step exactly with the help of the analytic solution in the Fourier domain; the nonlinear part is solved by the RK4 scheme.

In the test runs a planar Stokes wave with the steepness $\varepsilon = k_0H/2 = 0.3$ (waves break when $\varepsilon \approx 0.42$) defined in $N_x = 2^7$ grid points in the physical space and in a four times larger Fourier domain does not exhibit noticeable variation of the shape, when the nonlinear parameter is big, $M > 12$, and the time step is $\omega_0 dt = 2^{-7}$ (where the linear wave frequency is $\omega_0 = (gk_0)^{1/2}$) (Fig. 1a, for the first 7 periods the relative deviation in the amounts of the total energy is $1.5\cdot10^{-6}$, of the momentum is $1.9\cdot10^{-10}$); some variations of the high-order Fourier harmonics could be noticed when $\omega_0 dt = 2^{-5}$. However these runs of steep waves could not continue for longer than $O(10^1)$ wave periods and eventually stop due to the small-scale instability. The reduction of the nonlinear parameter of the scheme down to $M = 3$ leads to approximately constant values of only 4 first wave harmonics (Fig. 1b); as before, this simulation stops due to the unstable growth of small-scale Fourier components within a few wave periods (the relative errors in the total energy is $4.1\cdot10^{-6}$ and in the momentum is $5.0\cdot10^{-14}$ within the first 4 wave periods), even faster than when higher-order nonlinearity is taken into account. The small-scale instability could be damped using spectral filtering, though such filters were not used in the numerical experiments discussed in this paper.

The initial condition for the wave simulations is taken in the form of irregular waves which propagate in the direction of increasing coordinate $x$. Their Fourier amplitudes are specified in such a way that the linear solution would possess the JONSWAP frequency spectrum with the given peakedness $\gamma = 3$, significant wave height $4\sigma = 7$ m, peak wave period $T_p$ = 10 s and the characteristic angle $\theta$ = 62° of the $\cos^2$ directional spreading function $D(\chi)$,

$$D(\chi) = \begin{cases} \frac{2}{\theta}\cos^2\left(\frac{\pi\chi}{\theta}\right), & |\chi| \le \frac{\theta}{2} \\ 0, & |\chi| > \frac{\theta}{2} \end{cases}. \tag{5}$$

The initial condition at $t = 0$ was specified according to the linear theory in the spatial domain 50×50 dominant wave lengths (what is about 8 km by 8 km). The preparatory stage when the nonlinear terms are turning in force slowly (the nonlinear relaxation, following Dommermuth (2000)) occupies the first 20 dominant wave periods, $0 \le t < 200$ s, and then the waves are simulated for the following 20 min, 200 s $\le t <$ 1400 s. The calculated wave fields are stored each 0.5 s of the physical time. The collected data of the surface displacements are processed further. An example of the surfaces at the initial moment $t = 0$, at the time when the full nonlinearity is in force, $t$ = 200 s, and at the end of the simulation, $t$ = 1400 s, is given in Fig. 2. A qualitative difference between these three fields may be hardly discerned.

As mentioned above, the methods for wave breaking attenuation should affect the waves as little as possible. Two methods to regularize the wave breaking phenomenon are considered in this work. The first employs hyperviscosity causing dissipation at high wavenumbers. It is introduced with the help of extra terms, which act in the Fourier space, added to the evolution equations. Then the equations (1) and (2) may be represented in the Fourier domain in the following form,

$$\frac{\partial \hat{\eta}_{\vec{k}}}{\partial t} = F_1 - \alpha_k \hat{\eta}_{\vec{k}}, \qquad \frac{\partial \hat{\Phi}_{\vec{k}}}{\partial t} = F_2 - \alpha_k \hat{\Phi}_{\vec{k}}, \tag{6}$$

where hats denote spatial Fourier transforms and $F_1$ and $F_2$ are the Fourier transforms of the right-hand-sides in (1) and (2) correspondingly. Coefficients $\alpha_k$ are responsible for the viscosity at short scales, written in the form similar to Chalikov (2005),

$$\alpha_k = r k_{\max}\left[\frac{fk - k_{\max}}{(f-1)k_{\max}}\right]^p \quad \text{if} \quad \frac{k}{k_{\max}} \ge \frac{1}{f} \quad \text{and} \quad \alpha_k = 0 \quad \text{otherwise}, \tag{7}$$

$$k \equiv \sqrt{k_x^2 + k_y^2}\,.$$

Here $k_x$ and $k_y$ are the wave vector components. The number $k_{max}$ is the maximum wavenumber and $r > 0$, $p > 1$, $f > 1$ are coefficients. According to (7), the dissipation is absent ($\alpha_k = 0$) in the range of long waves $k \le k_{max} / f$, and it is maximum at the shortest scales ($\alpha_k = rk_{max}$ when $k = k_{max}$); $1/f$ is the fraction of the wavenumber space unaffected by the dissipative terms. The typical values of the parameters used for the present simulations are $r$ = 0.005, $f$ = 2, $p$ = 4. The power $p$ = 2 makes the additions to equations (1) and (2) similar to the terms of molecular viscosity.

The second method to avoid the wave breaking instability is a low-pass spectral filter similar to the one used in Xiao et al (2013). This filter is applied to the evolution equations at each iteration in time and may be written in the Fourier domain similar to (6) in the following form

$$\hat{\eta}_{\vec{k}} \to e^{-\beta_k}\hat{\eta}_{\vec{k}}, \qquad \hat{\Phi}_{\vec{k}} \to e^{-\beta_k}\hat{\Phi}_{\vec{k}}, \tag{8}$$

factors $\beta_k$ are functions of the wavenumber,

$$\beta_k = \left[ \frac{k}{m k_p} \right]^q . \qquad (9)$$

In (9) $k_p$ is the peak wavenumber, $m > 1$ is the inverse width of the mask in terms of $k_p$, and $q > 1$ is a parameter. In the simulations we typically use $m = 16$ and $q = 30$ similar to [Xiao et al, 2013]. It is obvious that $k_p$ in (9) may be straightforwardly related to $k_{max}$ in (7).

To illustrate how these approaches act, let us consider the solution of the evolution equations (1) and (2) with the help of a two-layer finite difference scheme in the Fourier domain. Then equations (6) may be written in the form

$$\hat{\eta}_{\vec{k}}(t+\Delta t) = \Delta t F_1(t) + \hat{\eta}_{\vec{k}}(t)(1-\Delta t \alpha_k), \quad \hat{\Phi}_{\vec{k}}(t+\Delta t) = \Delta t F_2(t) + \hat{\Phi}_{\vec{k}}(t)(1-\Delta t \alpha_k). \qquad (10)$$

Following the second approach (8), one iteration in time is described by the scheme

$$\hat{\eta}_{\vec{k}}(t+\Delta t) = \Delta t F_1(t) e^{-\beta_k} + \hat{\eta}_{\vec{k}}(t) e^{-\beta_k}, \quad \hat{\Phi}_{\vec{k}}(t+\Delta t) = \Delta t F_2(t) e^{-\beta_k} + \hat{\Phi}_{\vec{k}}(t) e^{-\beta_k}. \qquad (11)$$

It is instructive to expand $\exp(-\beta_k)$ into the Taylor series near $k = 0$ assuming $\beta_k \ll 1$,

$$e^{-\beta_k} \approx 1 - \beta_k, \qquad (12)$$

then (11) yields

$$\hat{\eta}_{\vec{k}}(t+\Delta t) \approx \Delta t F_1(t) + \hat{\eta}_{\vec{k}}(t)(1-\beta_k), \quad \hat{\Phi}_{\vec{k}}(t+\Delta t) \approx \Delta t F_2(t) + \hat{\Phi}_{\vec{k}}(t)(1-\beta_k), \qquad (13)$$

where the smallness of the time step $\Delta t$ has been utilized.

Note that despite much similarity between (10) and (13), they possess significant differences. At first, the application of a low-pass filter depends on the number of iterations in time (13), though the effect of hyperviscosity in (10) is proportional to $\Delta t$. It means that a choice of the time step effectively changes the effect of the low-pass filter. That is why the first approach (hyperviscosity) seems to be more physical. The filter (13) does not have a threshold minimum wavenumber, though the coefficient $\alpha_k$ (7) does. Furthermore, the power $q = 30$ is much greater than $p = 4$, which results in very strong dependence of the filter at large wavenumbers, see Fig. 3. Meanwhile the effective dissipation at large wavenumbers provided by these approaches seems to be comparable according to Fig. 3.

## 3. Conditions for stable energy preserving simulations of irregular waves

We assume the simulations of the potential hydrodynamic equations (1)-(4) by means of the HOSM to be the etalon reference framework. Fist, we would ensure that the result of simulation of the original Euler equations (i.e., conservative, with no breaking regularization) is accurate and stable with respect to the methods used for the discretization and integration in time. Since we are interested in the regimes when wave breaking occurs, the initial condition corresponds to relatively rough sea. The simulation of the Euler equations cannot be continued after the first breaking event happens as at this instant the Euler equations get inapplicable for description of the water surface. Numerically, this event causes failure of the Fourier decomposition associated with instability at short scales. Significantly, it is not obvious how to identify the events when the code blows up due to the physical breaking or due to a numerical instability (such as the examples in Fig. 1), therefore they both will be treated as breaking events. The strategy how to control the accuracy of the numerical simulation is not obvious either. The conservation of the physical conserved quantities, the mass, momentum and the total energy is checked to this end (see examples in Sec. 2). At the same time if one wave among the simulated 50×50 waves is totally removed from the field, it may result in no more than just 0.04% of the total energy loss. As will be seen later, a numerical simulation may be stable even having a significant change of the total energy. Furthermore, the shape of a particular wave may be greatly disfigured having moderate errors in conserving quantities; hence a fine conservation of the total energy actually does not guarantee that the large wave characteristics such as wave height and asymmetry are accurate.

The criterion of stability of the simulation with respect to variation of the resolution in time and space is used to obtain the trustworthy simulation. As was said above, the initial condition is the same in all the simulated cases. The results of various simulations have been thoroughly compared as described below. Parameters of the discussed simulations are given in Table 1. Two methods for the iterations it time were used, RK4 and A&RK4 (see Sec. 2). The second approach, A&RK4, when the linear part is calculated exactly, is usually considered to be more beneficial because the contribution of linear terms in the evolution equations is typically much bigger than the one of nonlinear terms. As a result a bigger time step may be used for simulations, and hence the calculation speeds up. Indeed, stable computations were possible with much larger time steps when the approach A&RK4 was used instead of RK4 (cf. Exp. 21 and Exp. 28 in Table 1).

However, from the results of a series of runs of the A&RK4 code with different time steps (Exp. 28, 29, 30, 39) we can conclude that the simulations are reproducible only when they have remarkably small time steps, what almost withdraws the advantage of the A&RK4 method. The simulation Exp. 28 with time step $dt = T_p/40$ is stable, while the simulations with sequentially smaller time steps blow up at $t = 717.5$ s. The simulation Exp. 39 ($dt = T_p/320$) is compared with Exp. 30, Exp. 29 and Exp. 28 in Fig. 4. In Fig. 4a the root-mean-square difference between the two simulated surfaces is plotted, which for the surfaces $\eta_a$ and $\eta_b$ is calculated following the formula

$$\Delta\eta = \sqrt{\frac{2\iint(\eta_a - \eta_b)^2\,dxdy}{\iint(\eta_a^2 + \eta_b^2)dxdy}}\,. \tag{14}$$

A similar characteristic is calculated in the Fourier domain (Fig. 4b),

$$\Delta\hat{\eta} = \sqrt{\frac{2\iint(|\hat{\eta}_a| - |\hat{\eta}_b|)^2\,dk_x dk_y}{\iint(|\hat{\eta}_a|^2 + |\hat{\eta}_b|^2)dk_x dk_y}}\,. \tag{15}$$

Note that the quantity (15) estimates the difference in absolute values of the Fourier amplitudes, it does not account for the disagreement in Fourier phases. Hence (15) characterizes the distribution of energy in the Fourier domain. We conclude from Fig. 4 that the simulations Exp. 28 and Exp. 29 are unacceptably inaccurate; Exp. 29 ($dt = T_p/80$) results in more than 10% disagreement in terms of $\Delta\eta$ within about 50 wave periods. The detailed investigation of Exp. 39 and Exp. 29 revealed that the maximum local difference between the surfaces was spread among many mesh points manifesting through short-scale wavy patterns.

The error $\Delta\eta$ in the simulation Exp. 30 compared to Exp. 39 grows gradually from zero to about 3% at the moment when the code blows up. The difference in the Fourier amplitudes makes about one half of $\Delta\eta$. The differences between the wave extremes found in Exp. 39 and Exp. 30 along the entire surfaces at given instants,

$$\delta_{\max} = 2\frac{\max_{(x,y)}(\eta_a) - \max_{(x,y)}(\eta_b)}{\max_{(x,y)}(\eta_a) + \max_{(x,y)}(\eta_b)}, \qquad \delta_{\min} = 2\frac{\min_{(x,y)}(\eta_a) - \min_{(x,y)}(\eta_b)}{\min_{(x,y)}(\eta_a) + \min_{(x,y)}(\eta_b)}, \tag{16}$$

have irregular dependence and at the final stage reach no more than 4%. The maximum local difference between the surfaces simulated in these runs was localized in one mesh point just before the break down of the simulation. In the simulation Exp. 30 the energy variation in the interval 200 s < $t$ < 700 s (50 wave periods) was within 0.001%. To sum up, we employ the time step $dt = T_p/160$ used in Exp. 30 as the standard time step for the following simulations, which provides reasonably accurate simulations of the water surfaces. In the literature from 32 (Xiao et al, 2013) to 200-500 (Toffoli & Bitner, 2014; Toffoli et al, 2008) steps per one wave period were used. Note that in the discarded Exp. 28 ($dt = T_p/40$) which clearly shows numerical artifacts the total energy is conserved within about 0.5%, what is similar to the accuracy of some stochastic simulations reported in the literature.

The comparison between the simulations with the use of RK4 and A&RK4 methods is provided in Fig. 5 for the runs Exp. 21 and Exp. 30. Both the simulations stopped after $t$ = 717.5 s due to the explosive instability. The root-mean-square difference between the two simulated surfaces (14) reaches about 4% (Fig. 5a). The difference between the Fourier amplitudes (15) is about twice smaller (Fig. 5b). The difference between the wave surface extremes (16) is no more than 5%.

When the integrations over the Fourier domain in (15) are performed in given intervals of the wavenumbers $k = (k_x^2 + k_y^2)^{1/2}$, the resulting function $\Delta\hat{\eta}(k)$ characterizes the difference associated with specific length scales; it is shown in Fig. 5c. The largest value of wavenumber in Fig. 5c corresponds to the maximum resolution in the lateral direction, $\max(k_y) \approx 5k_p$. Maximum of the discrepancy shown in Fig. 5c corresponds to large wavenumbers at large times, and it is rather small. It was found that the mismatch at short scales makes up the major part of the difference $\Delta\hat{\eta}$ plotted in Fig. 5b.

The wave profiles when the difference between the two simulated surfaces reaches the maximum are shown in Fig. 6 (only the interval of four dominant lengths along the $Ox$ axis is shown); this moment is just before the blowup of the simulation. The difference between the curves actually affects just a few adjacent mesh points; it is obviously related to the occurrence of a very steep surface slope with small length. Physically this process may be associated with microbreaking; it does not influence noticeably the dynamics of dominant waves, but causes numerical instability.

A similar agreement was observed between the simulations with twice finer resolution along the transverse direction, Exp. 53 and Exp. 24; the difference was even smaller in values. In these simulations the computations stopped even earlier, after $t$ = 375 s (about 17 periods in the fully nonlinear regime $t$ > 200 s). A spike develops in a very narrow area in the $Ox$ and $Oy$ directions similar to the previous case.

To conclude, for the selected parameters of simulations the methods RK4 and A&RK4 are almost equally valid and exhibit very similar results; still, a better stability of the simulation by A&RK4 may be expected.

The comparison between simulations with different spatial resolutions is limited by the circumstance that when the mesh is finer, the simulation blows up earlier (when spectral filters are absent). Thus, the simulations with the resolution 20×10 points per dominant wave length (Exp. 21 and Exp. 30) stopped after about 70 wave periods; the ones with the resolution 20×20 points (Exp. 24 and Exp. 53) – after less than 40 periods; and when we have 40×20 points – after less than a dozen periods from $t$ = 0, before the nonlinear terms turn in the full force (this simulation is not described in the paper).

The comparison between simulations Exp. 21 and Exp. 24 is performed for in total 37 periods of the evolution, 17 of them are in the regime of full nonlinearity, $t$ > 200 s. (A comparison between Exp. 30 and Exp. 53 give quite similar results.) The root-mean-square difference between the surfaces reaches almost 10%; about twice smaller difference is for the Fourier amplitudes. The maximum difference between the global surface extremes is up to about 5%. The maximum local difference between the surfaces is found well before the blowup of the higher-resolution simulation Exp. 24; it represents a wave-like small-scale pattern, as shown in Fig. 7, 8. The two left panels in Fig. 7 represent small areas in the vicinity of the point of the maximum difference (which is located at the centers of the panels). The panel to the right in Fig. 7 shows the difference between the surfaces. The corresponding longitudinal cuts in Fig. 8 have similar shapes, though the disturbance occupies many mesh points within about one wave length.

Though having some differences in Fig. 7 and 8 we have to regard that a finer resolution in space reduces the period of stable non-dissipative simulation. Hence we use the simulation Exp. 21 with the resolution 20×10 points per dominant wave length as the etalon non-dissipative numerical experiment (Exp. 30 could be used alternatively with the same outcomes as described in the next section). At the same time we simulate more stable codes with breaking regularization using a better resolution 20×20 points (i.e., the mesh $2^{10}\times 2^{10}$ for the domain 8×8 km). For the reference, in the study by Xiao et al (2013), where the low-pass filter was applied, the spatial resolution was about 32×32 points ($2^{12}\times 2^{12}$ for 128×128 dominant wave lengths). 28-56 grid points per one wave length were used in [Toffoli et al, 2008, 2010; Toffoli & Bitner-Gregersen, 2011]. In [Ducrozet et al, 2007, 2017] from 10 to 24 grid points discretized a wave in the longitudinal direction.

## 4. Simulations with hyperviscosity and with a low-pass filter versus the non-dissipating reference case

In this section we compare the results of the simulation of the same initial condition as used in Sec. 3, but when the wave breaking is handled with the help of hyperviscosity (6)&(7) or the low-pass spectral filter (8)&(9). Firstly we varied the parameter $r$ in (7) and $m$ in (9) to find the ones which ensure successful simulations till the time $t$ =1400 s and result in minimal possible dissipation (see Sec. 2 and Fig. 3).

The results of comparison of the reference conservative simulation (Exp. 21) versus the cases with hyperviscosity (Exp. 36, $r$ = 0.005, $f$ = 2, $p$ = 4) and with a low-pass spectral filter (Exp. 22, $m$ = 16, $q$ = 30) are given in Fig. 9. The total wave energy grows a little from $t$ = 0 to $t$ = 200 s when the preparatory nonlinear relaxation is going on, and then remains approximately constant in Exp. 21 until it rapidly increases at the moment of breakdown of the conservative simulation (Fig. 9a). The case with a low-pass filter (Exp. 22) preserves the energy with high accuracy, though the case with hyperviscosity (Exp. 36) exhibits some decay of the total energy. The difference between the simulated surfaces $\Delta\eta$ grows slowly from zero to about 40% (Fig. 9b). The difference between the Fourier amplitudes behaves similarly, though accounts for about half in the value (Fig. 9c). The distribution of the difference in the Fourier amplitudes in specific wavenumber bands is shown in Fig. 9d. The situations of the hyperviscosity and the low-pass filter cannot be distinguished in Figs. 9b-d.

The global extremes of the surfaces in Exp. 22 and Exp. 36 generally follow the reference simulation but not precisely. For times $t$ > 350 s significant mismatches between the extremes are observed in Exp. 21 vs Exp. 22, and Exp. 21 vs Exp. 36; they possess irregular character and can reach up to 20%. In Fig. 10 the surfaces in the vicinities of the maximum local differences are shown, simulated by the reference and the testee codes, and also the difference between the two surfaces. One may see that the wave surfaces simulated by the conservative and modified codes look quite similar, though the difference surfaces clearly exhibit wavy structures in both the cases. The locations shown in Fig. 10a,b and the instants when the maximum differences are observed are very close, but relatively far from the moment of breakdown of the non-dissipative simulation. The maximum differences between the surfaces are 7.28 m and 6.27 m correspondingly (having the significant wave height 7 m), and the root-mean-square differences are about 30% in both the cases.

The cuts of the surfaces in Fig. 10 along the direction of the wave propagation are shown in Fig. 11. The visible departure between the compared curves is localized within about one wave length; some noisy mismatches may be also noticed throughout the curves of the surface. Hence the seeming large value of discrepancy in terms of $\Delta\eta$ (Fig. 9b) in fact corresponds to quite localized mismatches which look insignificant in the wave slices.

## 5. Simulations by the codes using two different methods of wave breaking regularization

When the breaking is regularized, the simulation may be continued further beyond the time $t$ = 717.5 s when the non-dissipating simulation stops. The equations with hyperviscosity (Exp. 36) and with a low-pass filter (Exp. 22) were simulated further until $t$ = 1400 s; the results of these simulations are compared against each other in this section.

The wave energy variation for these two simulations may be found in Fig. 9a. The simulation with a low-pass filter (Exp. 22) does not exhibit visible variation for $t$ > 200 s (the energy loss is about 0.03%). In the case with hyperviscosity (Exp. 36) the total energy loss is more significant, about 2%. The difference between the simulated surfaces in the physical and Fourier domains are shown in Fig. 12. Similar to the previous cases, the excursion of Fourier phases makes about half of the mismatch. The maximum root-mean-square differences count about 60% and 30% respectively. The difference in the Fourier amplitudes in specific wavenumber bands is shown in Fig. 12c, it is mainly concentrated at small scales. A deeper analysis reveals that the wavenumbers in the range $k > 5k_p$ make similar contributions to the total disagreement of the Fourier amplitudes. Note that the maximum wavenumber limit in Fig. 12c is twice larger than in the previous figures as in these simulations $\max(k_x) = \max(k_y) \approx 10k_p$. Shapes of the wavenumber and frequency spectra of these two simulations generally coincide; they exhibit very small differences at the very ends of the tails. The differences between the extremes found along the entire surfaces at given instants are plotted in Fig. 13. Accumulating discrepancy is observed which results in the difference up to about 15%.

The surface patterns at the moments of large local discrepancies are opposed in Fig. 14. The maximum differences of 6.53 m (Fig. 14a) and 5.58 m (Fig. 14b) locate in the centers of the panels. The surfaces simulated by different models seem to be quite similar despite the root-mean-square difference exceeding 60% (hence, it is actually the case of a worst coincidence between the two simulations according to Fig. 12a). The differences (the rightmost panels in Fig. 14) are represented by noisy fragmented small-scale perturbations. The perturbations obviously may interfere and result in more noticeable transient local differences. The longitudinal sections of the surfaces in Fig. 14 along the points of the maximum difference are shown in Fig. 15. It follows from Fig. 15 that the non-systematic discrepancies between the simulated curves may indeed reach significant values, though they have characteristic scales much shorter than the dominant wave length, close to the mesh size.

## 6. Conclusion

In this work we focus on the issue of reliability of the results obtained within the direct numerical simulations of the evolution of 3D nonlinear surface sea waves. Focusing on the rogue wave problem, relatively rough sea states have been considered, which are inevitable affected by the wave breaking phenomenon. The wave overturning cannot be simulated by a simple model (while a heavy code cannot be used for the purpose of stochastic simulations), hence the wave breaking is parameterized based on phenomenological grounds. Ones may be faced with plenty of pitfalls on this way (such as an equivalence of the initial condition and the simulated sea state to the natural case; the effects of finite resolution in physical and Fourier domains; still idealized equations, etc). In this study we stress the point how two conventional methods to handle the wave breaking modify the solution on the typical scale of the sea wave quasistationarity ~20 min ($O(10^2)$ wave periods) in the dynamical and statistical sense.

In this study we employ the framework of the potential Euler equations for deep-water waves solved by the High Order Spectral Method and limited to the accurate accounting for four wave interactions. In the numerical simulations the dynamics of intense waves are affected by the wave breaking process associated with the short-scale instability in the Fourier domain, which cannot be properly described. Besides, the physical breaking is not reliably differentiated from the 'numerical breaking' due to the inner instability of the code. We chose the sea state parameters which may be simulated relatively long (for $O(10^1)$ wave periods), but wave breaking does occur in the wave fields within $O(10^2)$ wave periods. At first, we seek for the parameters when the simulations are reproducible using finer mesh and time steps. Compared to the numerous simulations by Xiao et al (2013), in our numerical experiments the spatial resolution is similar (20×20 points per dominant wave length), though the resolution in time is 5 times shorter (160 steps per dominant wave period), which was necessary to obtain a reproducible evolution of the water surface. In this simulation of the domain 8 km by 8 km (50 by 50 waves) the energy conservation error was within 0.001% for 50 wave periods. Rougher simulations could be formally even more stable than the accurate one, but exhibit noticeable differences in the wave dynamics.

Two conventional spectral approaches for the wave breaking regularization have been considered, using hyperviscosity and a spectral low-pass filter. The parameters of the methods were chosen to provide the weakest possible dissipation sufficient for stable simulations for 140 wave periods of the waves typical for the North Sea conditions. The considered wave breaking regularizations do influence the deterministic evolution of waves within 20 min: the difference in the instant extreme displacements between the simulations with and without the wave breaking regularization may reach up to 15%, the root-mean-square difference in surfaces is up to 60%. However, the difference between the surfaces calculated within the two regularized models is represented by small-scale noisy perturbations, which become appreciable at scales about one third of the dominant wave length and shorter. These noisy perturbations may form relatively large waves due to the random superposition. No preference to one of the considered approaches of the wave breaking regularization may be given based on the performed investigation. At the same time, the viscous terms lead to noticeably stronger decay of the total wave energy due to the slower power-law dependence in the Fourier domain.

When the simulations with regularized wave breaking are compared versus the reference non-dissipative case (prior to the first breaking event occurs), the discrepancy seems to be somewhat more noticeable. Contrary to the comparison between the two dissipative codes, the differences between the reference simulation with each of the regularized codes represent short-scale wave-like patterns, which visualize the effect of the wave breaking elimination. The discrepancy in the terms of the root-mean-square difference between the surfaces is also noticeably larger than the one between the two regularized codes.

The discussed effects are important for the deterministic description of sea waves and (at least partly) for the calculation of the wave spectra and wave statistics. Hence they should be taken into account when dealing with rare extreme events such as rogue waves as they increase the degree of uncertainty of the result of stochastic simulations, and when discussing the spectral energy cascades observed in direct numerical simulations.

## Acknowledgements

The support from the Russian Foundation for Basic Research (grants No. 17-05-00067 and 18-05-80019) is acknowledged by AK. AS is grateful for the support from the Fundamental Research Programme of RAS “Nonlinear Dynamics”.

**Table 1.** Parameters of the numerical simulations. The total energy difference is given for the interval 200 s < $t$ <1400 s.

| **Exp. No.** | **Breaking regularization** | **Method** | $N_x$ | $N_y$ | ***dt*, s** | **Accuracy** |
|---|---|---|---|---|---|---|
| **28** | no | A&RK4 | $2^{10}$ | $2^{9}$ | $2^{-2}$ | energy increase $4.9 \cdot 10^{-3}$ |
| **29** | no | A&RK4 | $2^{10}$ | $2^{9}$ | $2^{-3}$ | early stop at $t$ = 717.5 s |
| **30** | no | A&RK4 | $2^{10}$ | $2^{9}$ | $2^{-4}$ | early stop at $t$ = 717.5 s |
| **39** | no | A&RK4 | $2^{10}$ | $2^{9}$ | $2^{-5}$ | early stop at $t$ = 717.5 s |
| **53** | no | A&RK4 | $2^{10}$ | $2^{10}$ | $2^{-4}$ | early stop at $t$ = 375.5 s |
| **21** | no | RK4 | $2^{10}$ | $2^{9}$ | $2^{-4}$ | early stop at $t$ = 717.5 s |
| **24** | no | RK4 | $2^{10}$ | $2^{10}$ | $2^{-4}$ | early stop at $t$ = 375 s |
| **22** | low-pass filter | RK4 | $2^{10}$ | $2^{10}$ | $2^{-4}$ | energy loss $2.7 \cdot 10^{-4}$ |
| **36** | hyper viscosity | RK4 | $2^{10}$ | $2^{10}$ | $2^{-4}$ | energy loss $1.9 \cdot 10^{-2}$ |

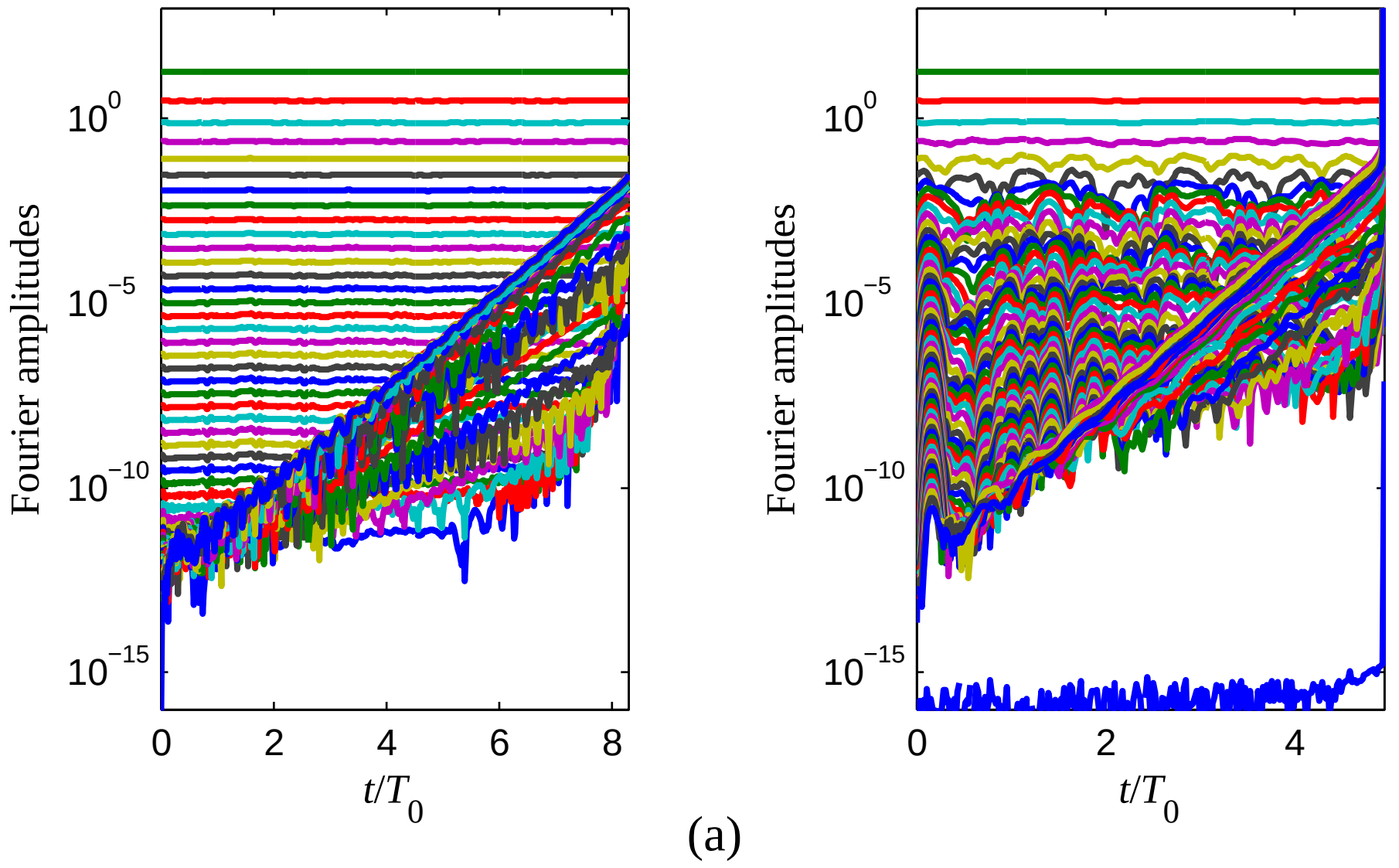

(a) (b)

**Fig. 1.** Evolution of the Fourier amplitudes of the planar Stokes wave with the steepness $k_0H/2 = 0.3$, resolution $N_x = 2^7$, $\omega_0 dt = 2^{-7}$ and parameters $M = 15$ (a) and $M = 3$ (b). No spectral filter applied to attenuate the small-scale instability. $T_0 = 2\pi/\omega_0$ is the linear wave period.

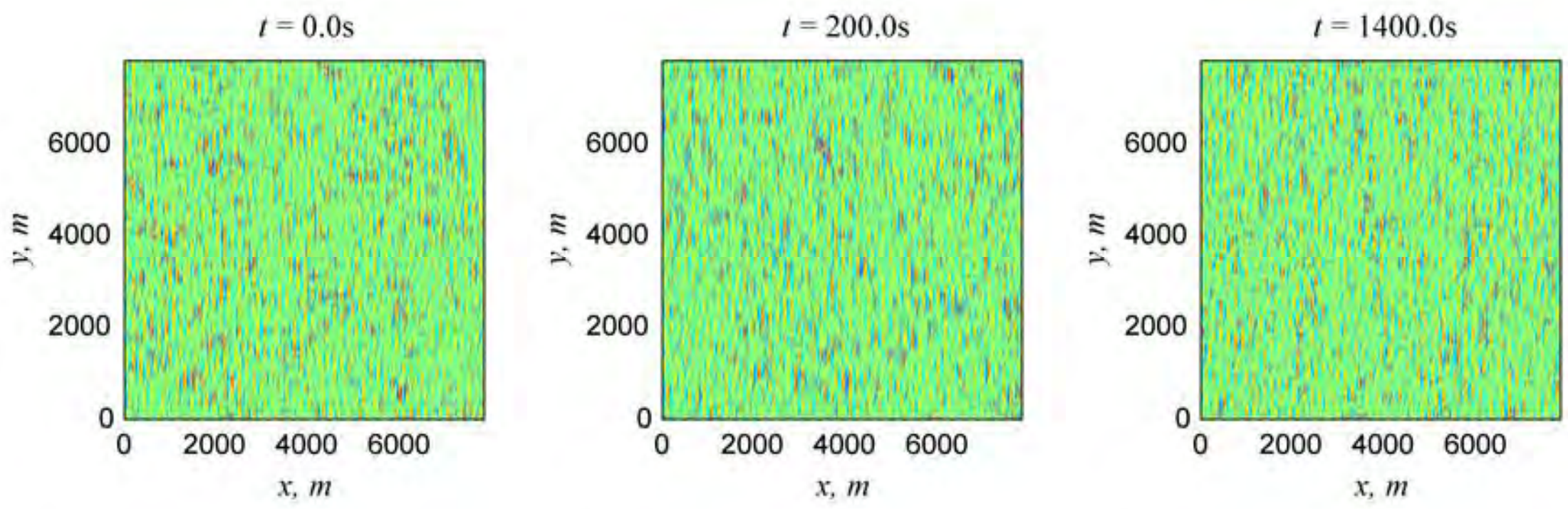

**Fig. 2.** Example of the simulated sea surfaces (Exp. 36).

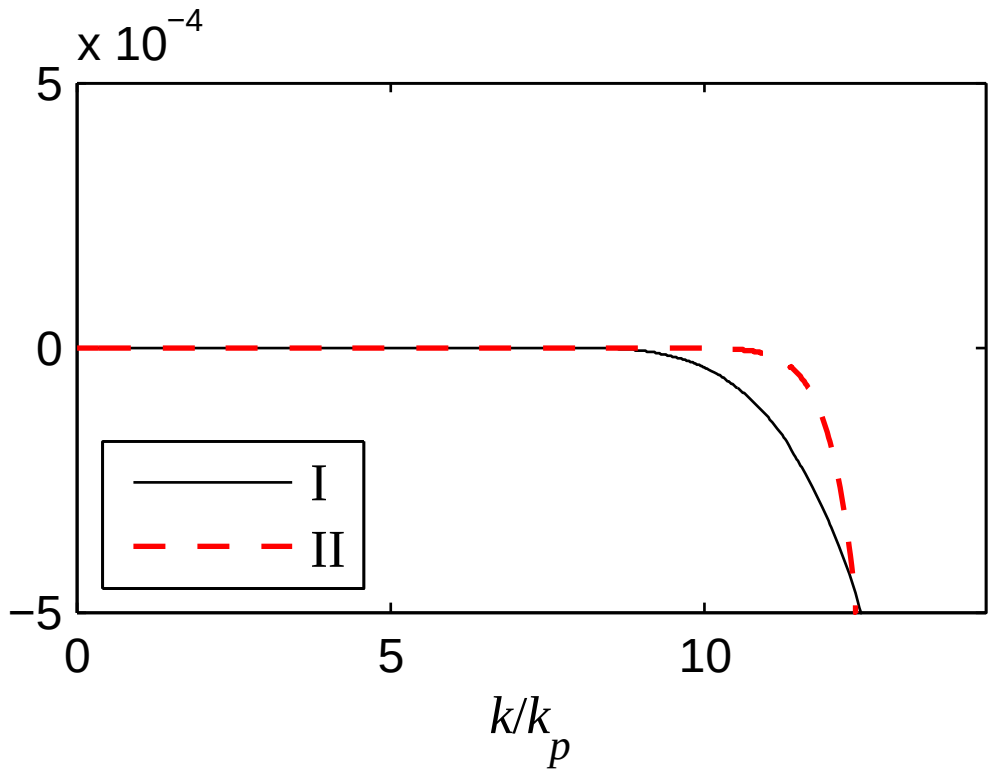

**Fig. 3.** Comparison of the functions $-\Delta t\alpha_k$ (I) and $(\exp(-\beta_k) - 1)$ (II) for the typical conditions of the numerical simulations. The parameters are $r = 0.05$, $p = 4$, $f = 2$, $m = 16$ and $q = 30$ (Exp. 36 and Exp. 22 respectively); $k_{max} \approx 10k_p$.

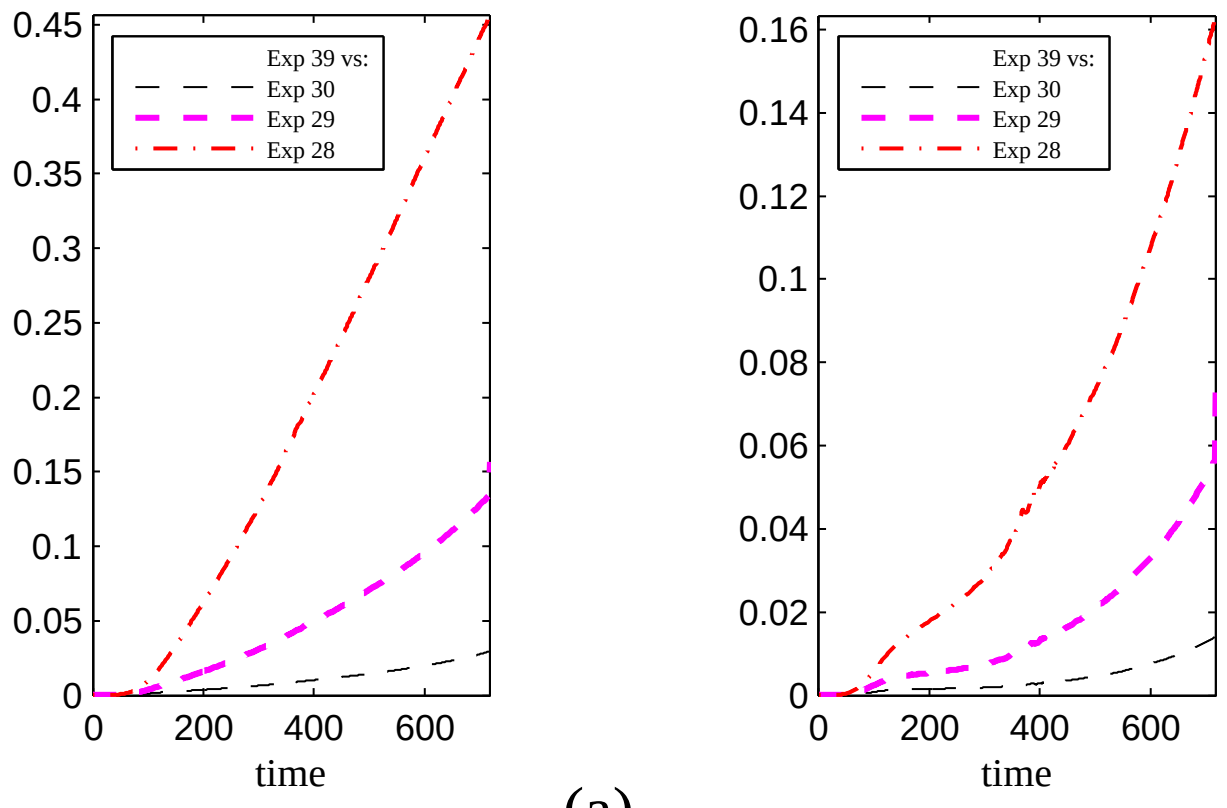

(a) (b)

**Fig. 4.** Comparison between the simulations with different time steps, Exp. 39 versus Exp. 30, Exp. 29 and Exp. 28: the difference (14) between the surface displacements (a) and the difference (15) between the Fourier amplitudes for surface displacements (b).

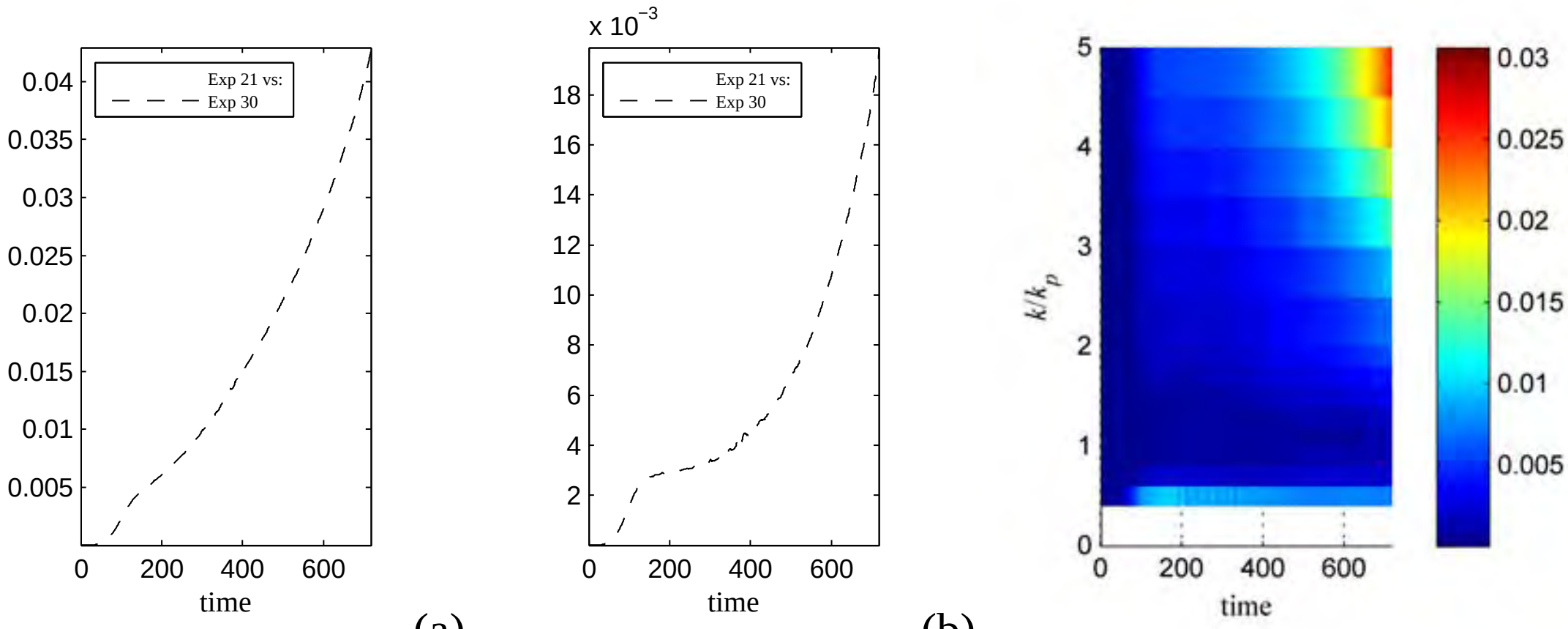

(a) (b) (c)

**Fig. 5.** Comparison between the simulations using the RK4 and A&RK4 methods (Exp. 21 and Exp. 30 respectively): the difference between the surface displacements (a); the difference between the Fourier amplitudes for surface displacements (b); the difference in specific wavenumber bands (c).

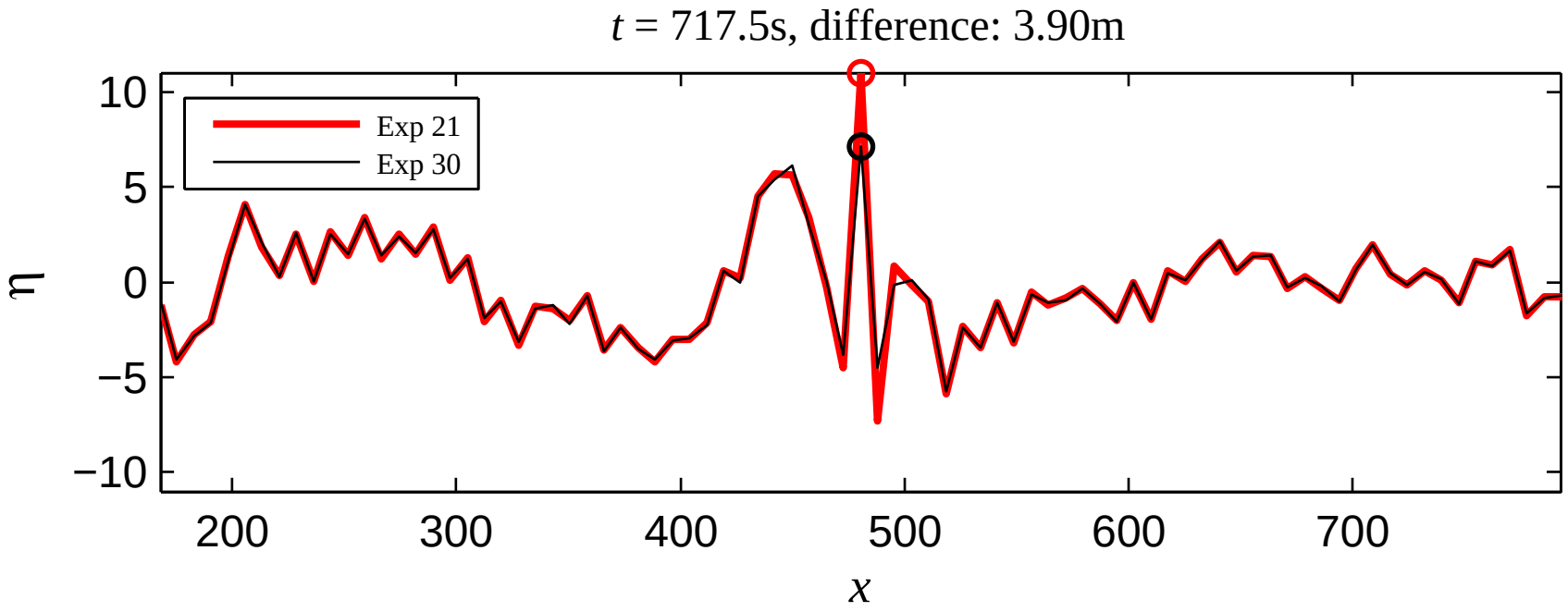

**Fig. 6.** The longitudinal cuts of the surfaces at the moment of maximum difference between the surfaces along the point of the maximum difference (shown with circles). The shown domain is of the size of 4 dominant wave lengths.

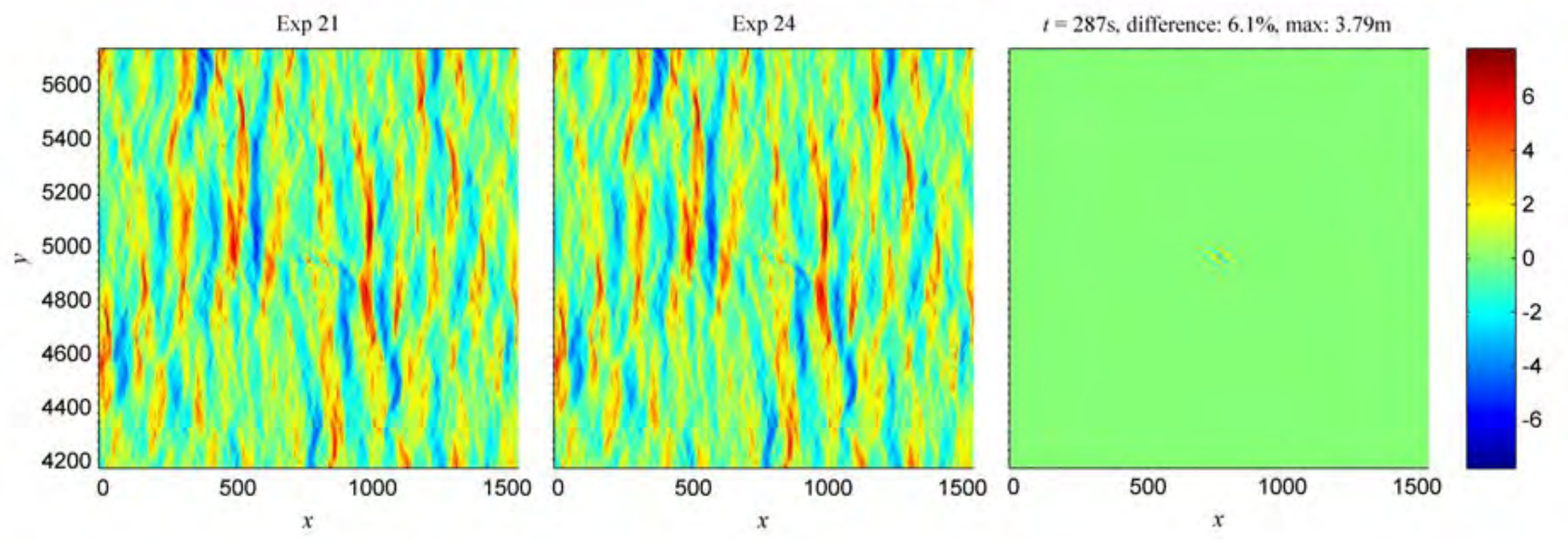

**Fig. 7.** The instants when the maximum local difference between the simulated surfaces of different resolution is achieved (Exp. 21 vs Exp. 24). The areas of the size 10×10 dominant wave lengths (two left panels) and the corresponding difference (the rightmost panels) are shown. The time instants, $\Delta\eta$ and the maximum difference are listed above the panel from the right.

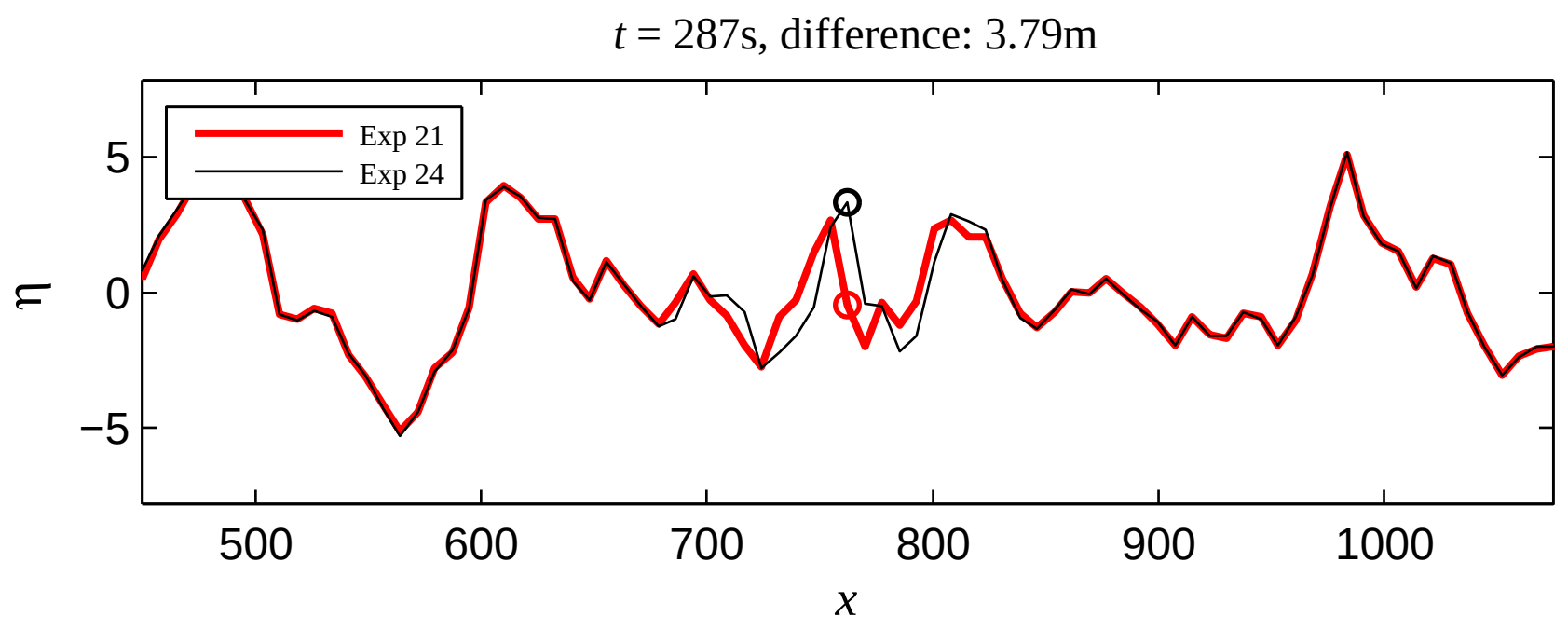

**Fig. 8.** The longitudinal cuts of the surfaces at the moment of maximum difference between the surfaces shown in Fig. 7 along the point of the maximum difference (shown with circles).

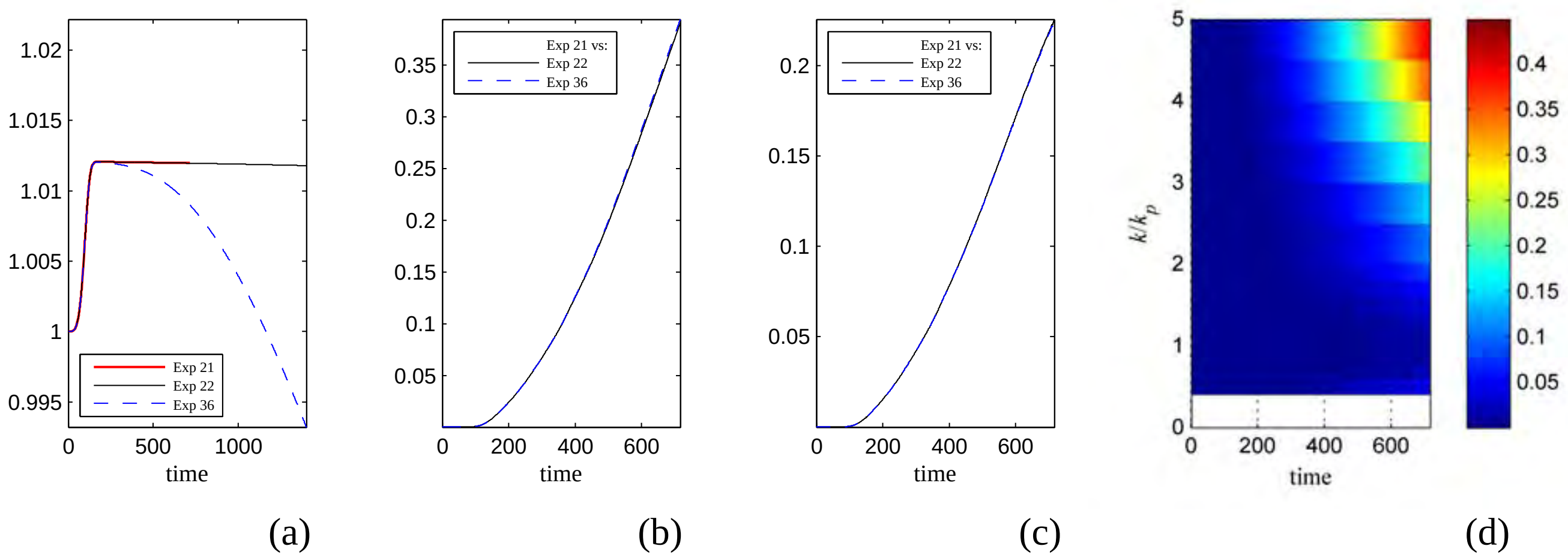

**Fig. 9.** Comparison of the non-dissipative simulation (Exp. 21) versus the one with weak hyperviscosity (Exp. 36) and the other with a low-pass spectral filter (Exp. 22): variation of the total energy with respect to the initial value (a); the difference between the surface displacements (b); the difference between the Fourier amplitudes for surface displacements (c); the difference in specific wavenumber bands between the surfaces in Exp. 21 and Exp. 36 (d).

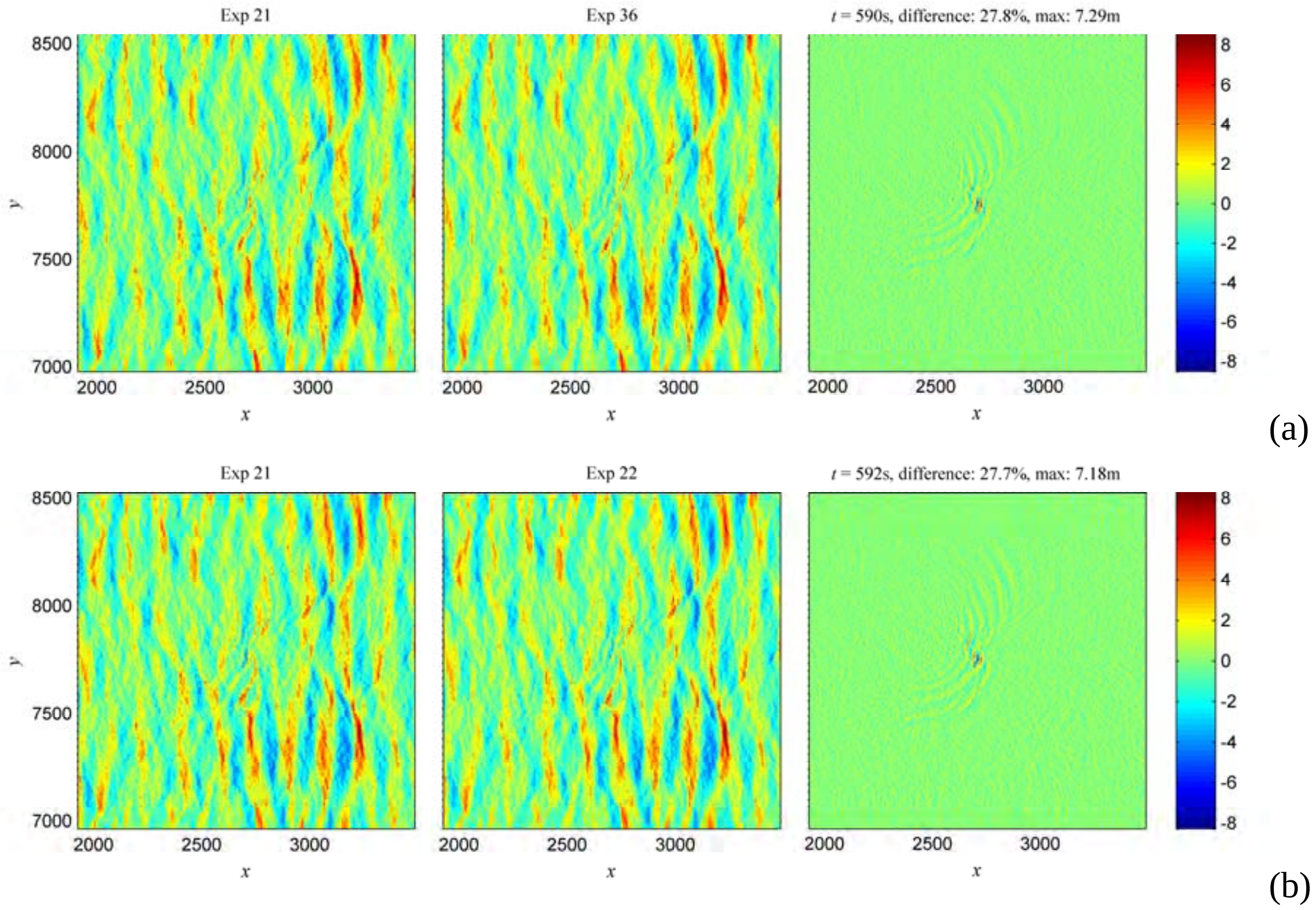

**Fig. 10.** The surfaces at the instants when the maximum local differences are achieved, and their differences: Exp. 21 versus Exp. 36 (a) and Exp. 21 versus Exp. 22 (b).

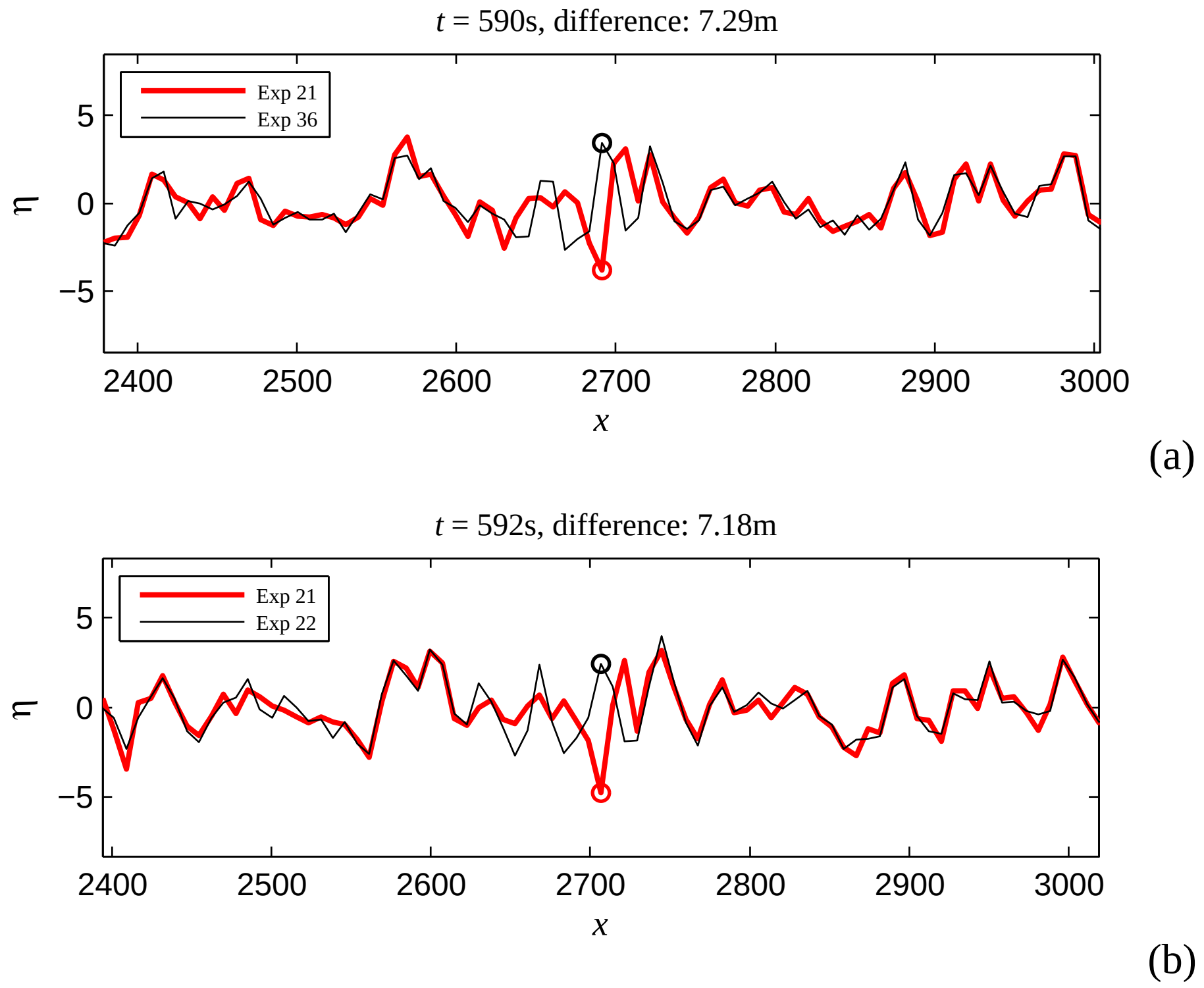

**Fig. 11.** The longitudinal cuts of the surfaces shown in Fig. 10 along the points of the maximum differences between the surfaces (shown with circles).

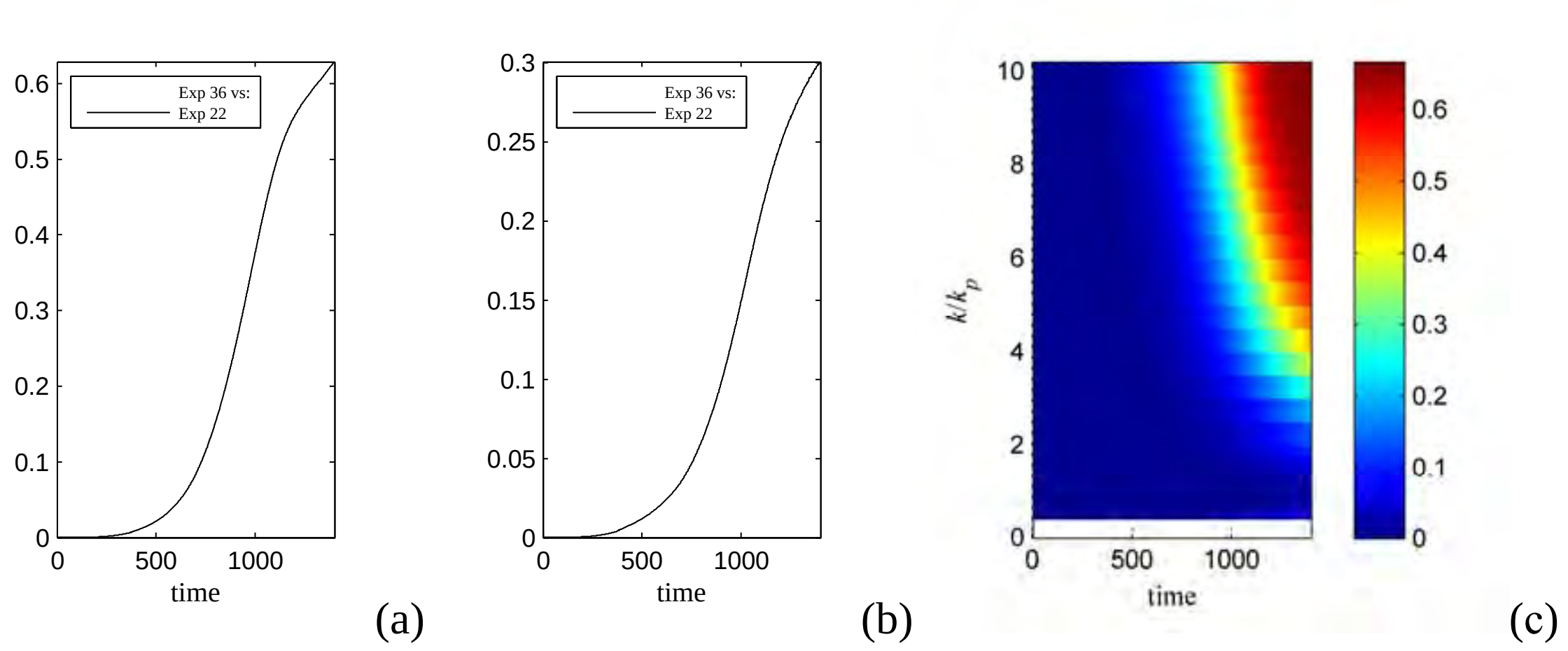

**Fig. 12.** Comparison between the simulation with weak hyperviscosity (Exp 36) and with a low-pass spectral filter (Exp 22): the difference between the surface displacements (a); the difference between the Fourier amplitudes for surface displacements (b); the difference in specific wavenumber bands (c).

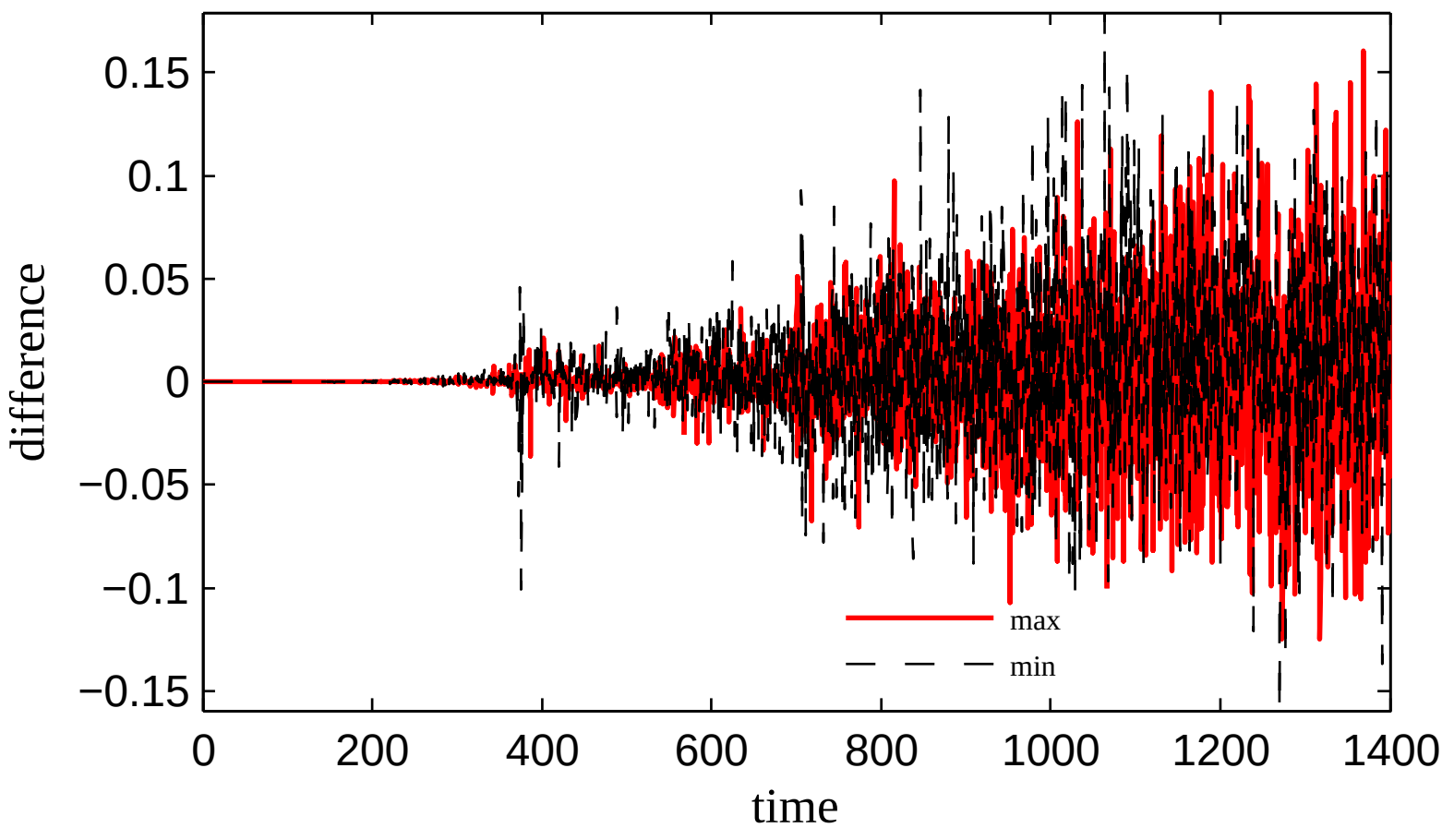

**Fig. 13.** The evolution in time of the differences between the global over the surface wave maxima (solid line) and the global minima (dashed line) observed in the Exp. 22 and Exp. 36.

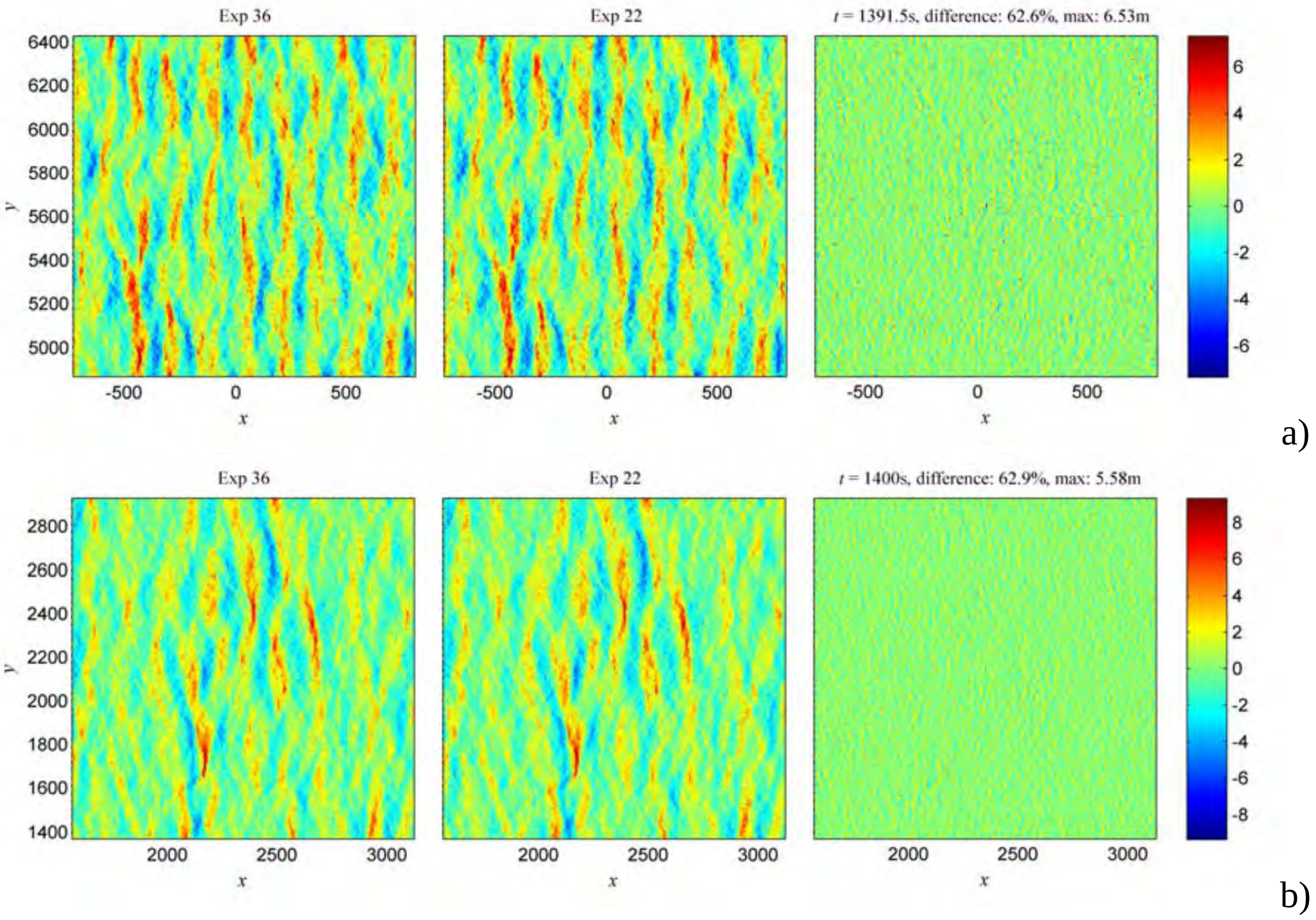

**Fig. 14.** Two instants when large local differences between the surfaces simulated in Exp. 36 and Exp. 22 were achieved.

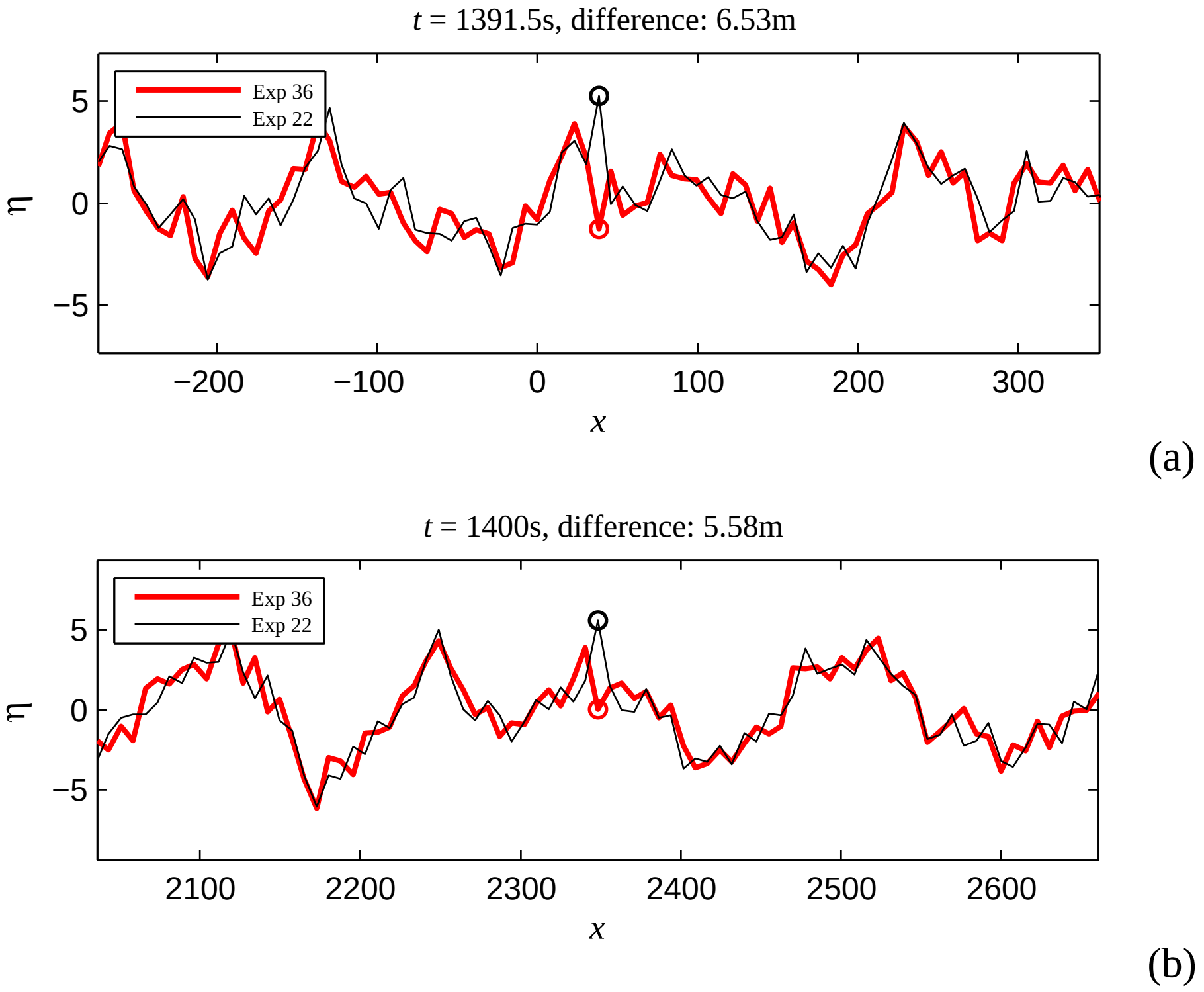

**Fig. 15.** The longitudinal cuts of the surfaces shown in Fig. 14 along the points of the maximum difference between the surfaces (shown with circles).